\input{aipcheck}

\documentclass[
    ,final            
  ]
  {aipproc}

\layoutstyle{8x11single}

\begin{document}

\title{Background analysis and status of the ANAIS dark matter project}

\author{J. Amar\'{e}, S. Cebri\'{a}n, C. Cuesta\footnote{Present address: Department of Physics, Center for Experimental Nuclear Physics and Astrophysics, University of Washington, Seattle, WA, USA}, E. Garc\'{i}a, C. Ginestra, M. Mart\'{i}nez\footnote{Present address: Universit\`{a} di Roma La Sapienza, Piazzale Aldo Moro 5, 00185 Roma, Italy}, M. A. Oliv\'{a}n, Y. Ortigoza, A. Ortiz de Sol\'{o}rzano, C. Pobes\footnote{Present address: Instituto de Ciencia de Materiales de Arag\'{o}n, Universidad de Zaragoza-CSIC, Calle Pedro Cerbuna 12, 50009 Zaragoza, Spain}, J. Puimed\'{o}n, M. L. Sarsa, J. A. Villar, and P. Villar}{  address={Laboratorio de F\'{i}sica Nuclear y Astropart\'{i}culas, Universidad de Zaragoza, Calle Pedro Cerbuna 12, 50009 Zaragoza, Spain\\
Laboratorio Subterr\'{a}neo de Canfranc, Paseo de los Ayerbe s/n, 22880 Canfranc Estaci\'{o}n, Huesca, Spain}
}

\begin{abstract}
ANAIS (Annual modulation with NaI Scintillators) is a project aiming to set up at the new facilities of the Canfranc Underground Laboratory (LSC), a large scale NaI(Tl) experiment in order to explore the DAMA/LIBRA annual modulation positive result using the same target and technique. Two 12.5\,kg each NaI(Tl) crystals provided by Alpha Spectra took data at the LSC in the ANAIS-25 set-up. The comparison of the background model for the ANAIS-25 prototypes with the experimental results is presented.  ANAIS crystal radiopurity goals have been achieved for $^{232}$Th and $^{238}$U chains, but a $^{210}$Pb contamination out-of-equilibrium was identified, whose origin has been studied. The high light collection efficiency obtained with these prototypes allows to anticipate an energy threshold of the order of 1 keVee. A new detector, with improved performances, was received in March 2015 and very preliminary results are shown.
\end{abstract}

\maketitle


\section{The ANAIS experiment}

The ANAIS (Annual modulation with NaI Scintillators) project is intended to search for dark matter annual modulation with ultrapure NaI(Tl) scintillators at the Canfranc Underground Laboratory (LSC) in Spain. The motivation of the ANAIS experiment is to provide a model-independent confirmation of the annual modulation positive signal reported by the DAMA collaboration~\cite{DAMAphaseI} using the same target and technique. To achieve this goal, ANAIS detectors should have similar performance to DAMA/LIBRA ones in terms or threshold and background, but as we will report below, reducing the threshold below 2\,keVee will imply a good sensitivity to the DAMA/LIBRA singled out region in the WIMP parameter space even in the case of higher background.

The total NaI(Tl) active mass will be divided into modules, each consisting of a 12.5\,kg NaI(Tl) crystal encapsulated in copper and optically coupled to two photomultipliers (PMTs) working in coincidence. Several modules, accounting for around 100\, kg, will be set-up at LSC along next months. The shielding for the experiment consists of 10\,cm of archaeological lead, 20\,cm of low activity lead, 40\,cm of neutron moderator, an anti-radon box (to be continuously flushed with boil-off nitrogen), and an active muon veto system made up of plastic scintillators designed to cover top and sides of the whole ANAIS set-up. The hut that will house the experiment at the hall B of LSC (under 2450\,m.w.e.) is already operative, shielding materials and electronic chain components are prepared for mounting. Different PMT models were tested in order to choose the best option in terms of light collection and background. The Hamamatsu R12669SEL2 was selected, and all the units are available at the LSC.

The main challenge has been the achievement of the required low background, in particular the development of crystals having a potassium content at the ppb level. A modular approach has been followed, from the ANAIS-0 module to the ANAIS-25 and ANAIS-37 set-ups. The ANAIS-0 module, which consists in a 9.6\,kg ultrapure NaI(Tl) crystal, 4"\,x\,4"\,x\,10", made by Saint-Gobain was first operated. It was thoroughly studied to characterize ANAIS background, optimize events selection, design the calibration method, test the acquisition code and electronics, and determine the optimum configuration of PMTs and light guides. The main results have been published in~\cite{ANAISbulk,ANAISbkg,anais40K,ANAISom}. Two prototypes of 12.5\,kg mass (named D0 and D1), made by Alpha Spectra with ultrapure NaI powder took data at the LSC from December 2012 to March 2015 for a general performance and background assessment. We will refer in the following to this set-up as ANAIS-25. The ANAIS-37 set-up combines the ANAIS-25 modules with a new module also built from Alpha Spectra, using improved protocols for purification and growing of the powder and crystal, designed in view of the ANAIS-25 results.

\section{ANAIS-25}

The main goals for this set-up were to measure the crystal internal contamination, determine light collection efficiency, fine tune the data acquisition and test the filtering and analysis protocols. ANAIS-25 set-up consisted of two modules of 12.5\,kg each provided by Alpha Spectra, Inc. Colorado.  The modules are cylindrical, 4.75$"$ diameter and 11.75$"$ length, with quartz windows for PMTs coupling. A Mylar window in the lateral face allows for low energy calibration. Two types of photomultiplier were tested: One module coupled to two Hamamatsu  R12669SEL2 and the other coupled to Hamamatsu R11065SEL, later replaced by the  R12669SEL2 model too. The modules were surrounded by 10\,cm of archaeological lead plus 20\,cm of low activity lead shielding at the Canfranc Underground Laboratory. ANAIS-25 modules took data from December 2012 to March 2015.

The first feature to be remarked is the excellent light collection as it can be seen in Table~\ref{tab:ASphekev}, especially with the Ham. R12669SEL2 which is the PMT model chosen to be used at the ANAIS experiment. This light collection has a good impact in both resolution and energy threshold.

\begin{table}
\begin{tabular}{lll}
\hline
  \tablehead{1}{l}{b}{Detector}
  & \tablehead{1}{l}{b}{PMT model}
  & \tablehead{1}{l}{b}{phe./keV}\\
\hline
D0 & Ham. R12669SEL2		&15.6\,$\pm$\,0.2\\
D1 & Ham. R11065SEL		   &12.6\,$\pm$\,0.1\\
D1 & Ham. R12669SEL2		&15.2\,$\pm$\,0.1\\
\hline
\end{tabular}
\caption{Light collection efficiencies (phe./keV) for ANAIS-25 detectors, derived from the 22.6\,keV line ($^{109}$Cd calibration).}
\label{tab:ASphekev}
\end{table}

Background contributions have been thoroughly analyzed. A detailed study of cosmogenic radionuclide production in NaI(Tl) derived from these data can be found at references~\cite{ANAIScosmo,ANAIScosmoLRT}. The preliminary results of ANAIS-25 were published in~\cite{ANAISricap13} and Table~\ref{tab:A25cont} shows the results of the activities determined for the main crystal contaminations: $^{40}$K content has been measured performing coincidence analysis between 1461\,keV and 3.2\,keV energy depositions in different detectors~\cite{anais40K} and the activities from $^{210}$Pb and $^{232}$Th and $^{238}$U chains have been determined on the one hand, by quantifying Bi/Po sequences, and on the other, by comparing the total alpha rate with the low energy depositions attributable to $^{210}$Pb, which are fully compatible. These results give a moderate contamination of $^{40}$K, above the initial goal of ANAIS (20\,ppb of K) but acceptable, a high suppression of $^{232}$Th and $^{238}$U chains but a high activity of $^{210}$Pb at the mBq/kg level. The origin of such contamination was identified and has been addressed by Alpha Spectra (see next section).

\begin{table}
\begin{tabular}{llll}
\hline
  \tablehead{1}{l}{b}{$^{40}$K (mBq/kg)}
  & \tablehead{1}{l}{b}{$^{238}$U (mBq/kg)}
    & \tablehead{1}{l}{b}{$^{210}$Pb (mBq/kg)}
  & \tablehead{1}{l}{b}{$^{232}$Th (mBq/kg)}\\
\hline
1.25\,$\pm$\,0.11 (41\,ppb K)& 0.010\,$\pm$\,0.002& 3.15 &0.0020\,$\pm$\,0.0008\\

\hline
\end{tabular}
\caption{Internal contamination measured in the ANAIS-25 crystals.}
\label{tab:A25cont}
\end{table}

\begin{figure}
 \includegraphics[width=0.45\textwidth]{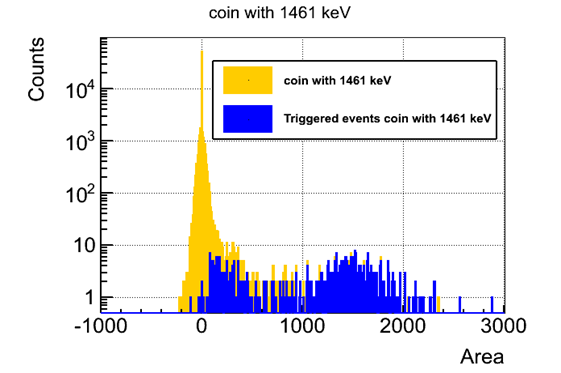}
  \includegraphics[width=0.45\textwidth]{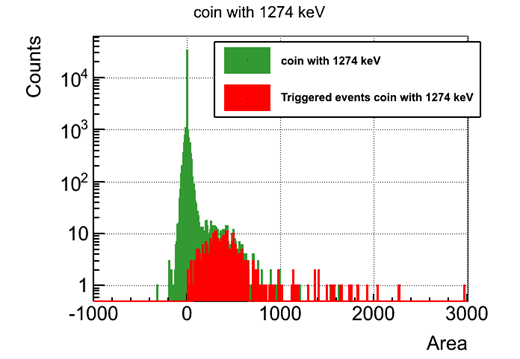}

  \caption{ANAIS-25 D0 coincident events at low energy for $^{40}$K (left), and for $^{22}$Na (right).}
  \label{fig:A25trigger}
\end{figure}

\begin{figure}
 \includegraphics[width=0.5\textwidth]{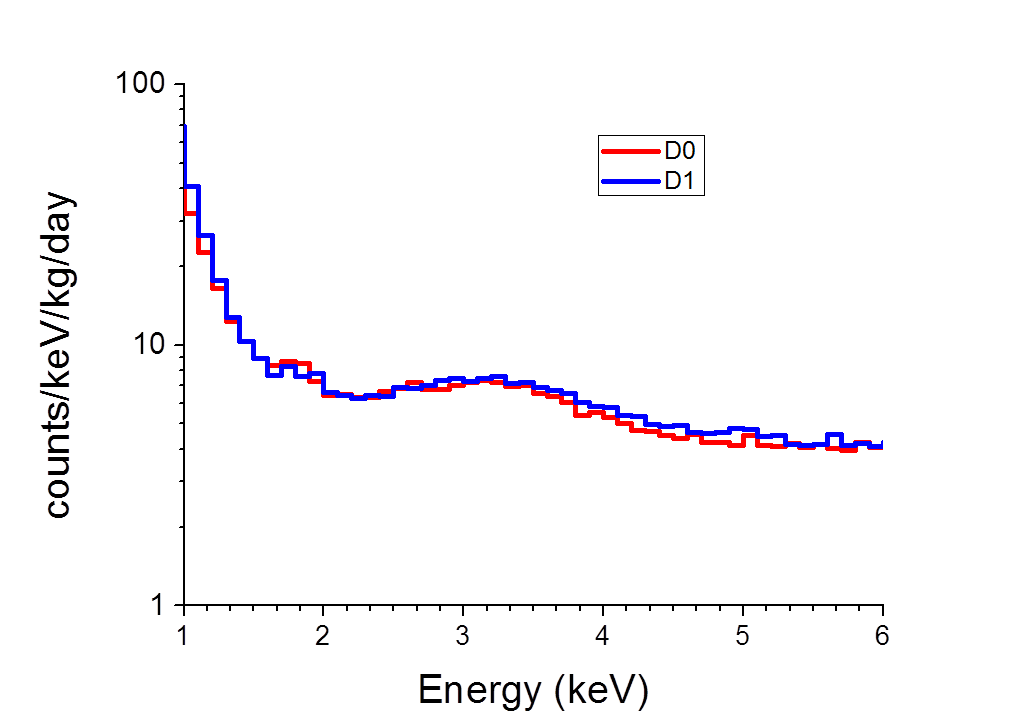}
  \caption{Preliminary filtered background spectra corrected by triggering and filtering efficiencies for ANAIS-25 detectors D0 and D1 (filtering procedures still being optimized).}
  \label{fig:A25le}
\end{figure}

A preliminary study has been carried out by using low energy events populations from internal $^{40}$K and $^{22}$Na. The K-shell electron binding energy following electron capture in $^{40}$K (3.2\,keV) and $^{22}$Na (0.9 keV) can be tagged by the coincidence with a high energy $\gamma$ ray (1461\,keV and 1274\,keV respectively). In Figure~\ref{fig:A25trigger} are shown both populations tagged by the high energy gamma, together with the events effectively triggering our acquisition. From Figure~\ref{fig:A25trigger}, it can be concluded that triggering at 1\,keVee is clearly achieved in ANAIS-25 detectors and then, an energy threshold of the order of 1\,keVee is at reach. The main issue to reach the 1\,keVee threshold is the effective removal of PMT origin events, which are dominating the background below 10\,keVee. Following the work done in~\cite{ANAISbulk} we have designed specific filtering protocols for ANAIS-25 detectors, being triggering and filtering efficiencies still under study. A preliminary spectrum, after filtering and correcting by the efficiencies of the cuts, determined with low energy events from a Cd-109 calibration, is shown in Figure~\ref{fig:A25le}.

\begin{figure}
\includegraphics[height=0.3\textheight]{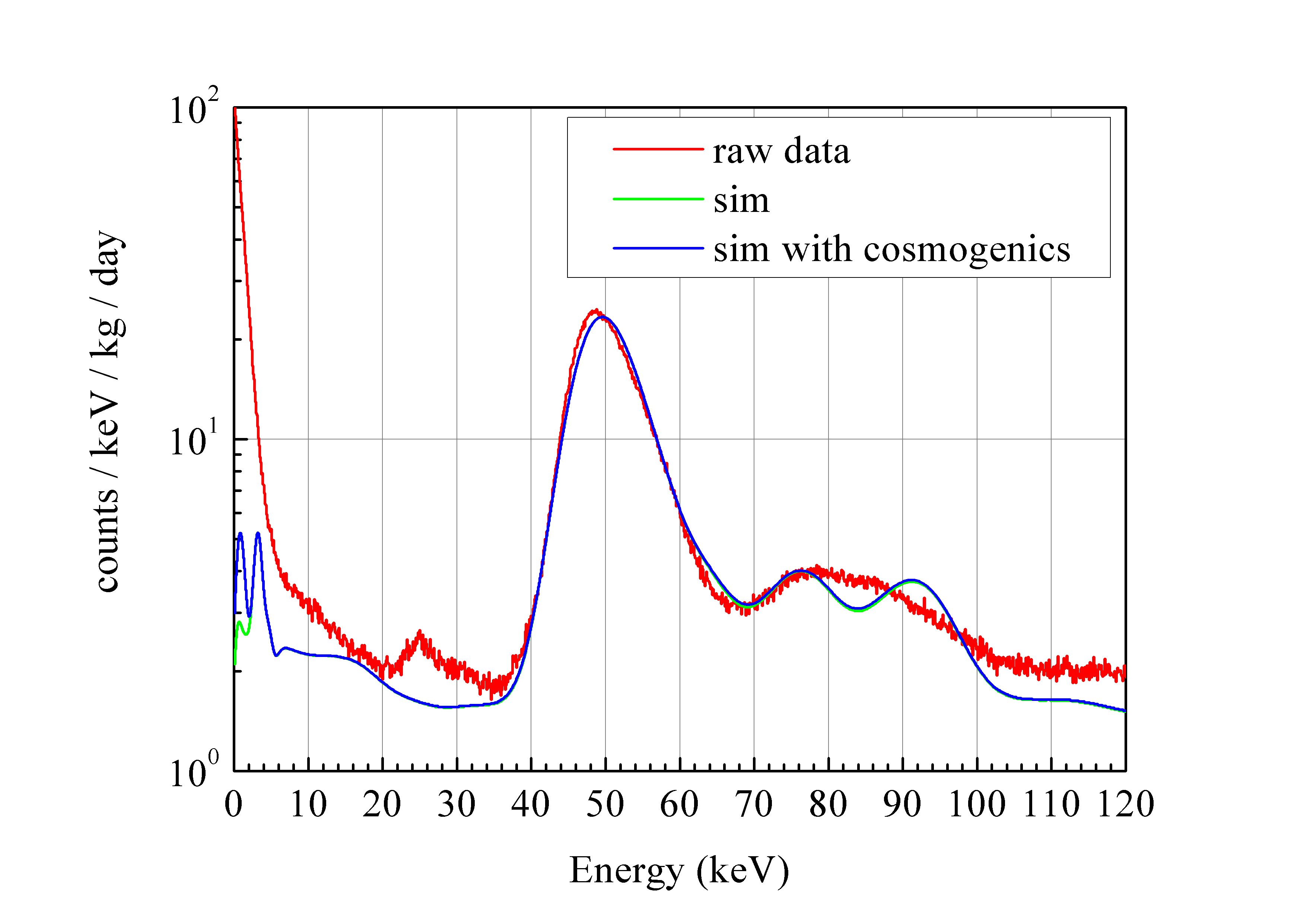}
\includegraphics[height=0.3\textheight]{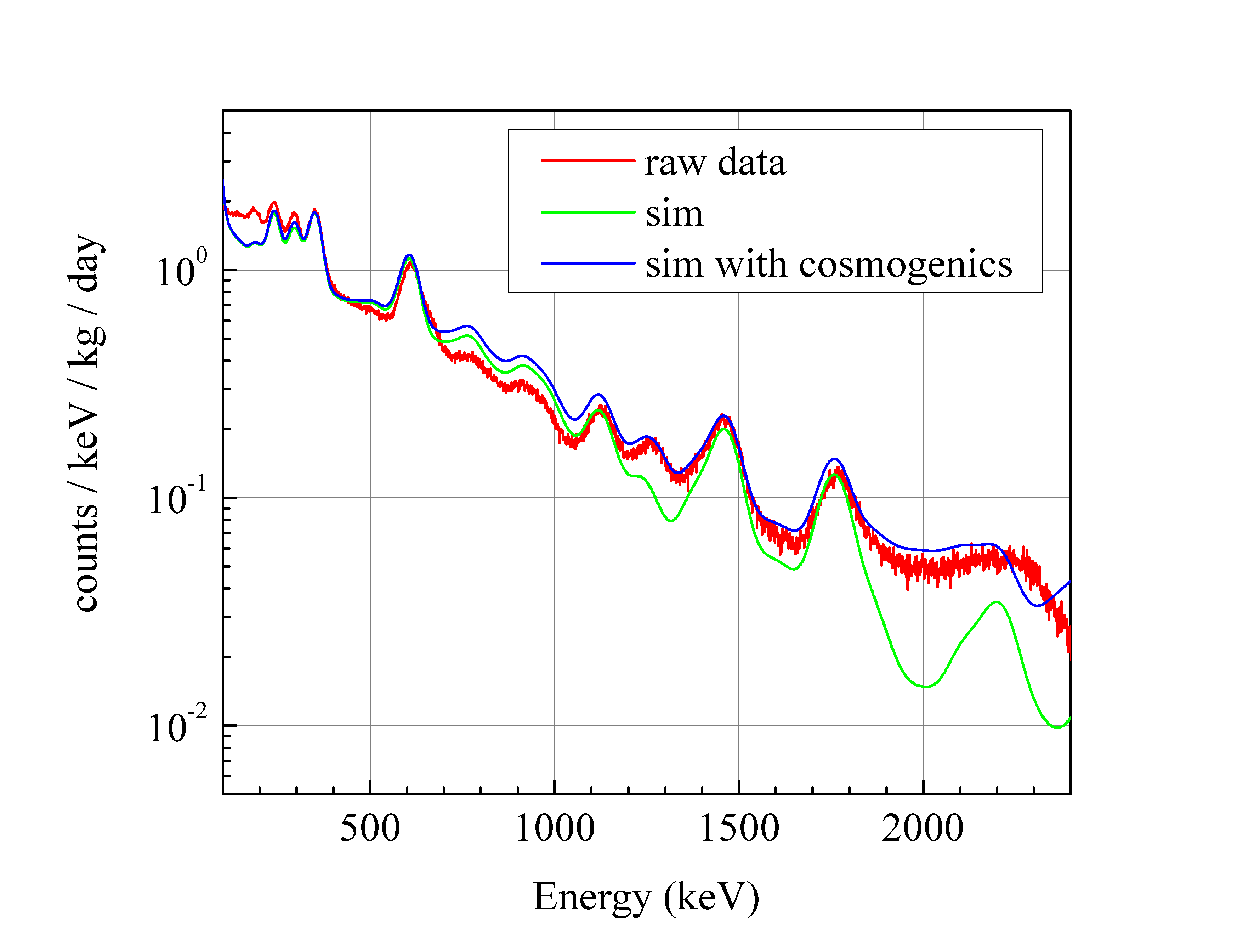}
 \caption{Comparison of the energy spectra summing all the simulated contributions (before and after adding cosmogenics) with the measured raw data for ANAIS-25 D0, considering anticoincidence data. Two different energy ranges are shown.}
 \label{fig:A25sim}
\end{figure}

\begin{figure}
\includegraphics[width=0.55\textwidth]{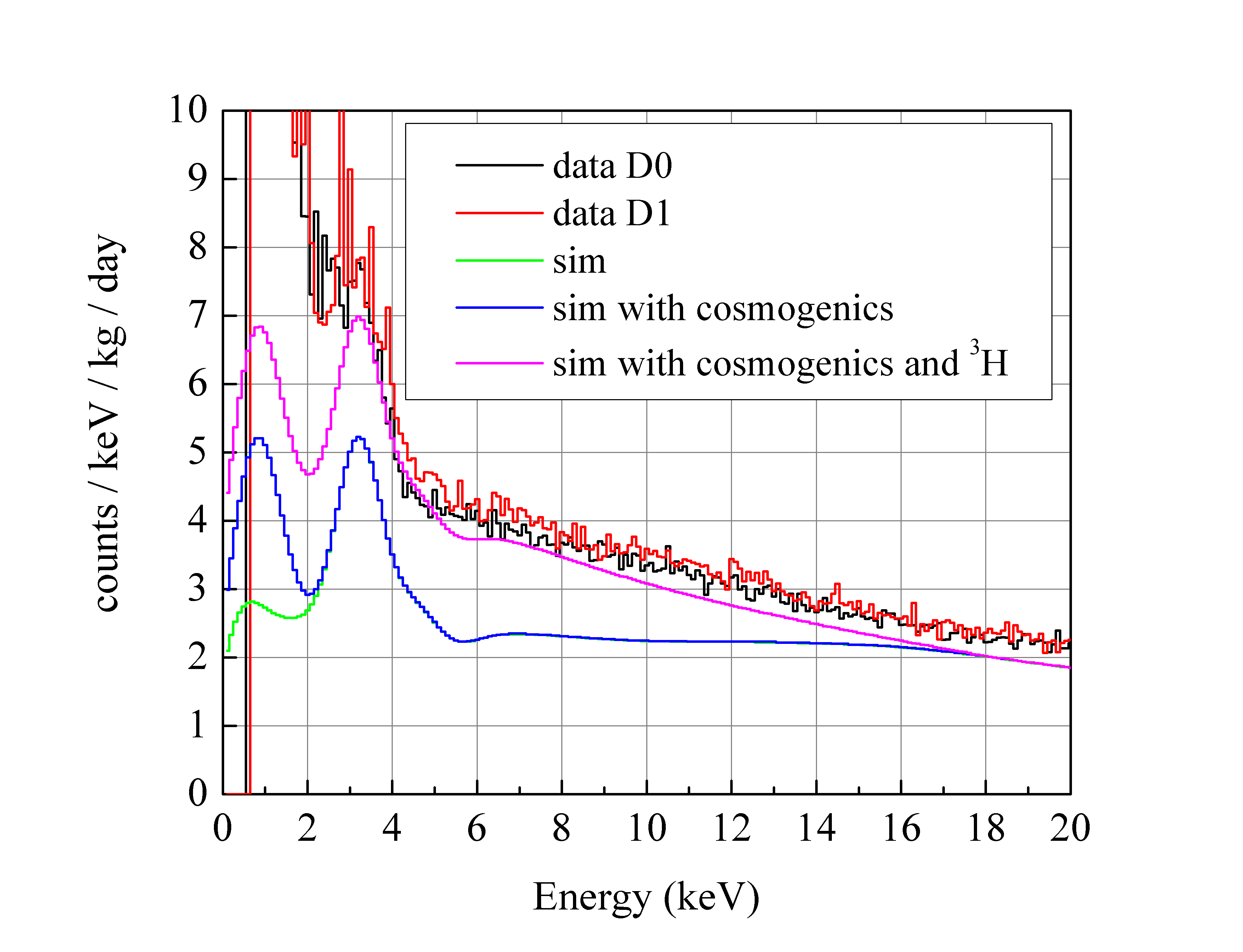}
 \caption{Comparison of the very low energy spectra of simulations with the measured data for ANAIS-25 detectors, , after filtering and considering anticoincidence, including $^{3}$H in the model.}
 \label{fig:A25sim2}
\end{figure}

The background model of the ANAIS-25 modules has been developed following the same procedure than for the previous prototype ANAIS-0~\cite{ANAISbkg}. The background sources considered in the model include: activities from external components such as PMTs, copper encapsulation, quartz windows, silicone pads, and archaeological lead; contribution from radon of the air filling the inner volume of the shielding, as one hundredth of the measured external air Rn activity since the volume inside the shielding is flushed with boil-off nitrogen; intrinsic activities from the NaI(Tl) crystals (as reported on Table~\ref{tab:A25cont}) and concerning $^{129}$I, since there is a broad range of activity values in iodine compounds (depending on the ore origin) and a direct quantification was not possible, concentration of the isotope was assumed to be the same as estimated by DAMA/LIBRA ($^{129}$I/$^{nat}$I\,=\,(1.7\,$\pm$\,0.1)$\cdot$10$^{13}$)~\cite{DAMAapparatus}; and cosmogenic origin activities in the NaI(Tl) crystals as quantified in~\cite{ANAIScosmo}. The contribution of these background sources has been assessed by Monte Carlo simulation
using the Geant4 code. Figure~\ref{fig:A25sim} compares the energy spectra summing all the simulated contributions described above with the measured data for ANAIS-25 detectors, considering anticoincidence data. A good agreement is obtained at high energy, but in the very low energy region some contribution seemed to be missing. Since upper limits on radionuclide activity have been used for several components, the background seems to be overestimated in some energy regions. Inclusion of cosmogenics was essential to reproduce coincidence data. It was found that the inclusion in the model of an additional activity of $\sim$0.2\,mBq/kg of $^{3}$H in the NaI crystals significantly improves the agreement with data at low energy, as shown in Figure~\ref{fig:A25sim2}. This value is about twice the upper limit set for DAMA/LIBRA crystals ($<$0.09\,mBq/kg~\cite{DAMAapparatus}), but lower than the saturation activity which can be deduced from the production rate at sea level of $^{3}$H in NaI, calculated in~\cite{mei} as explained in~\cite{ANAIScosmoLRT}. Figure~\ref{fig:A25sim2} summarizes the different contributions from the explained background model of ANAIS-25 detectors, for anticoincidence data, to the rate in the region from 1 to 10\,keV.

\section{ANAIS-37}

\begin{figure}
 \includegraphics[width=0.7\textwidth]{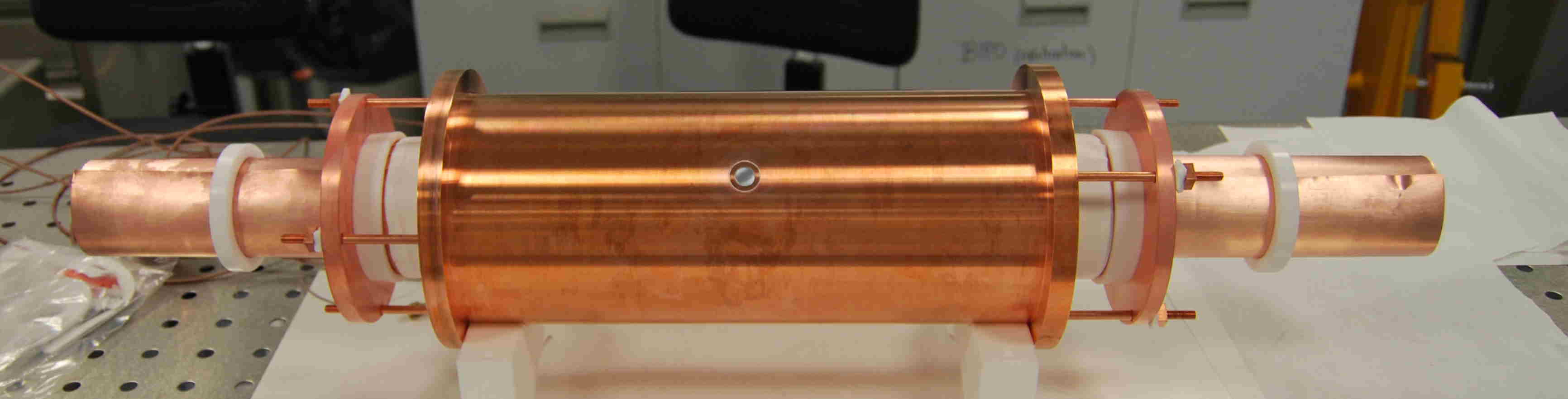}
  \caption{New NaI(Tl) module that is part of the ANAIS-37 set-up.}
  \label{fig:A37}
\end{figure}

\begin{figure}
 \includegraphics[height=0.25\textheight]{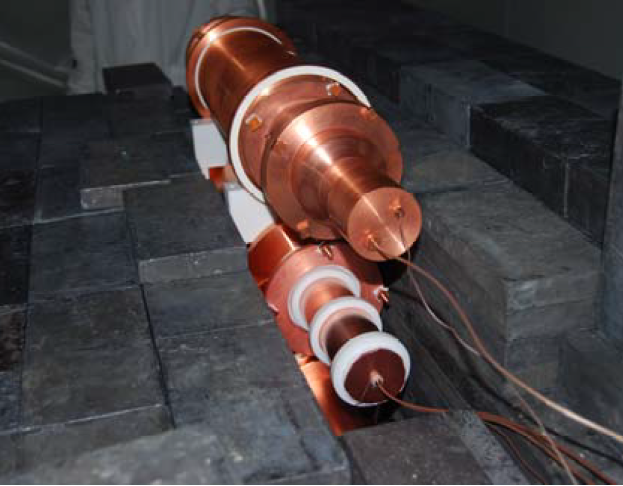}
 \includegraphics[height=0.25\textheight]{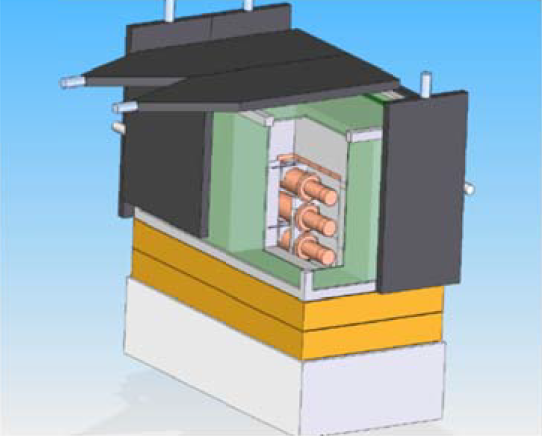}
  \caption{Left, picture taken during installation of ANAIS-37 set-up. Right, schematic drawing of the ANAIS-37 experimental layout at LSC consisting of 10\,cm archaeological lead plus 20\,cm low activity lead, all enclosed in a PVC box continuously flushed with boil-off nitrogen and active vetoes anti-muons.}
  \label{fig:A37setup}
\end{figure}

Alpha Spectra new module consists in a 12.5\,kg crystal made with a more purified powder, grown under improved conditions in order to prevent radon contamination, according to Alpha Spectra information. The crystal was encapsulated following similar protocols and using same materials to those required for ANAIS-25 detectors. An aluminized Mylar window was also built to allow low energy calibrations of the module. The crystal was received on the 6$^{th}$ of March, 2015 and Ham. R12669SEL2 PMTs were coupled to this new module at LSC clean room (see Figure~\ref{fig:A37}). Profiting from the ANAIS-25 setup, this new crystal was very fast commissioned and data taking started the 11$^{th}$ of March, 2015. We will refer in the following to this set-up as ANAIS-37, and it consists in three modules, 12.5\,kg mass each. The new module (D2) is placed in between the two ANAIS-25 modules (D0 and D1) to maximize the coincidence efficiency for the potassium determination (see Figure~\ref{fig:A37setup}).

Very preliminary results corresponding to 50 days of live-time are presented here. The total alpha rate of the new module D2 is determined through pulse shape analysis. A corresponding alpha rate of 0.58\,$\pm$\,0.01\,mBq/kg has been observed, which is a factor 5 lower than alpha rate in ANAIS-25 modules (3.15\,mBq/kg). We can conclude that effective reduction of Rn entrance in the growing and/or purification at Alpha Spectra has been achieved. 

The potassium content of the new D2 crystal has been analyzed using the same technique applied to previous prototypes. Bulk $^{40}$K content is estimated by searching for the coincidences between 3.2\,keV energy deposition in one detector (following EC) and the 1461\,keV gamma line escaping from it and being fully absorbed in the other detector. Efficiency of the coincidence was estimated using Geant4. We can conclude that the new D2 crystal has a potassium content of 44\,$\pm$\,4\,ppb compatible with that obtained with previous Alpha Spectra crystals (shown in Table~\ref{tab:A25cont}).

The light collection efficiency has been estimated following the same procedure as in the previous modules. As a result, 15.4\,$\pm$\,0.1\,phe./keV are obtained with this new module, also compatible with that obtained with previous Alpha Spectra modules and the Ham. R12669SEL2 PMTs (shown in Table~\ref{tab:ASphekev}).

\section{Sensitivity}

\begin{figure}
 \includegraphics[width=0.5\textwidth]{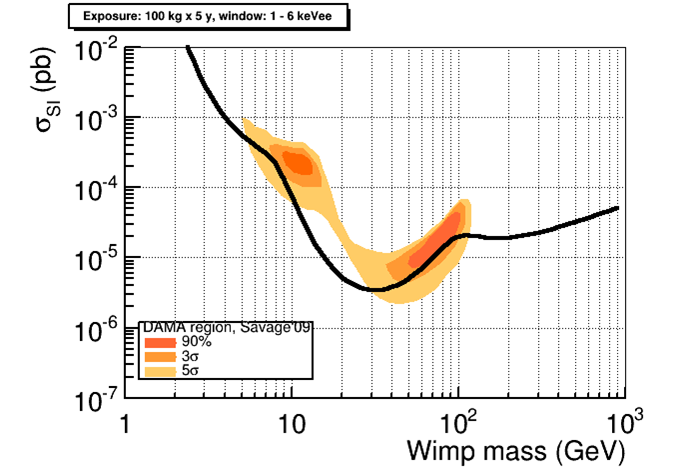}
  \caption{Prospects of the sensitivity to the annual modulation for 100\,kg total detection mass and presently achieved background (module D2) in the framework of the ANAIS experiment. Five years data taking have been assumed and an energy window from 1 to 6\,keVee. These prospects correspond to a detection limit at 90\% CL with a critical limit at 90\% CL, according to sensitivity estimates proposed in~\cite{sensitivityPlots}.}
  \label{fig:sens}
\end{figure}

Prospects of the sensitivity to the annual modulation in the WIMP mass - cross-section parameter space are shown in Figure~\ref{fig:sens} for 100\,kg configuration and 5 years of data taking. The analysis window considered is from 1 to 6\,keVee. The background assumed is the one measured in ANAIS-25 (corrected by the efficiencies of the cuts applied to remove PMT events and the trigger efficiency, and shown in Figure~\ref{fig:A25le}), but the $^{210}$Pb activity measured in the new module D2, i.e. the contribution of 2.57\,mBq/kg of $^{210}$Pb has been subtracted to the background measured at ANAIS-25. Further reduction from anticoincidence measurement, dependent on the matrix assumed, is expected. The most conservative approach to derive these prospects has been followed, but even in this case there is a considerable discovery potential of dark matter particles as responsible of the DAMA/LIBRA signal. An energy threshold lower than 2\,keVee is a crucial issue in this moment for the ANAIS experiment in order to guarantee enough sensitivity to test the DAMA/LIBRA result.

\begin{theacknowledgments}
  This work was supported by the Spanish Ministerio de Econom\'{i}a y Competitividad and the European Regional Development Fund (MINECO-FEDER) (FPA2011-23749), the Consolider-Ingenio 2010 Programme under grants MULTIDARK CSD2009-00064 and CPAN CSD2007-00042, and the Gobierno de Arag\'{o}n (Group in Nuclear and Astroparticle Physics, ARAID Foundation and C.~Cuesta predoctoral grant). C.~Ginestra and P.~Villar are supported by the MINECO Subprograma de Formaci\'{o}n de Personal Investigador. We also acknowledge LSC and GIFNA staff for their support.
\end{theacknowledgments}



\bibliographystyle{aipproc}   

\bibliography{LRT15_CCuesta_ANAIS}

\IfFileExists{\jobname.bbl}{}
 {\typeout{}
  \typeout{******************************************}
  \typeout{** Please run "bibtex \jobname" to optain}
  \typeout{** the bibliography and then re-run LaTeX}
  \typeout{** twice to fix the references!}
  \typeout{******************************************}
  \typeout{}
 }

\end{document}